 \documentclass[pmlr,twocolumn]{jmlr} 

\usepackage{booktabs}
\usepackage[load-configurations=version-1]{siunitx} 

\theorembodyfont{\upshape}
\theoremheaderfont{\scshape}
\theorempostheader{:}
\theoremsep{\newline}

\jmlryear{2021}
\jmlrworkshop{AAAI Workshop on Diversity in Artificial Intelligence (AIDBEI 2021)} 

\usepackage{rotating}
\usepackage{tablefootnote}

\title[Measuring Diversity of Artificial Intelligence Conferences]{Measuring Diversity of Artificial Intelligence Conferences}

\author{%
\Name{Ana Freire} \Email{ana.freire@upf.edu}\\
\Name{Lorenzo Porcaro} \Email{lorenzo.porcaro@upf.edu}\\
\addr Universitat Pompeu Fabra, Barcelona\\
    Roc Boronat, 138. 08018 Barcelona (Spain)
\AND
\Name{Emilia Gómez} \Email{emilia.gomez@upf.edu}\\
\addr Joint Research Centre, European Commission. \\
Edificio Expo, Calle Inca Garcilaso, 3. 41092 Sevilla (Spain)
}

\begin{document}

\maketitle

\begin{abstract}
The lack of diversity of the Artificial Intelligence (AI) field is nowadays a concern, and several initiatives such as funding schemes and mentoring programs have been designed to overcome it. However, there is no indication on how these initiatives actually impact AI diversity in the short and long term. This work studies the concept of diversity in this particular context and proposes a small set of diversity indicators (i.e. indexes) of AI scientific events. These indicators are designed to  quantify  the  diversity  of  the  AI  field  and  monitor  its  evolution.  We  consider  diversity  in  terms  of  gender, geographical location and business (understood as the presence of academia versus industry). We compute these indicators for the different communities of a conference: authors, keynote speakers and organizing committee. From these components we  compute a  summarized  diversity  indicator  for  each  AI event. We evaluate the proposed indexes for a set of recent major AI conferences and we discuss their values and limitations. 
\end{abstract}

\section{Introduction}

It is well recognized that Artificial Intelligence (AI) field is facing a diversity crisis, and that the lack of diversity contributes to perpetuate historical biases and power imbalance.
Different reports, such as the European Ethics guidelines for trustworthy AI\footnote{\url{https://ec.europa.eu/digital-single-market/en/news/ethics-guidelines-trustworthy-ai}} and the last AI Now Institute report (\cite{west2019discriminating}), emphasize the urgency of fighting for diversity and re-considering diversity in a broader sense, including gender, culture, origin and other attributes such as discipline or domain that can contribute to a more diverse research and development of AI systems. 

As a consequence, the research community has established different initiatives for increasing diversity such as mentoring programs, visibility efforts, travel grants, committee diversity chairs and special workshops\footnote{See, for instance, the activities launched by the Women in Machine Learning initiative: \url{https://wimlworkshop.org}}. However, there is no mechanism to measure and monitor the diversity of a scientific community and be able to assess the impact of these different initiatives and policies. 

In order to address that we propose a methodology to monitor the diversity of a scientific community. We focus on a set of international, well-recognized scientific conferences as they are the most relevant outcome at the moment for AI research dissemination. We consider diversity in terms of gender, geographical location and academia vs industry (possibly to extend further) and incorporate three different aspects of a scientific conference: authors, keynote speakers and organizers. After a literature review on diversity, we present the proposed indicators and illustrate them in a set of impact AI conferences.

\section{Background} \label{background}
\subsection{The concept and measurement of diversity}
Addressing the problem of conceptualizing diversity is a long-lasting debate in the academic community, object of study of several disciplines such as ecology, geography, psychology, linguistics, sociology, economics and communication, among the others. 
The interest in estimating the degree of diversity is often justified by the relevance of its possible impact: from the promotion of pluralism and gender, racial and cultural equality, to the enhancement of productivity, innovation and creativity in sociotechnical systems (\cite{Stirling2007}). 
Introducing its ubiquity, in a very broad sense Stirling defines diversity as "\textit{an attribute of any system whose elements may be apportioned into categories}". 
%


We can find in the previous definition two words which reflect different dimensions of diversity: \textit{elements} and \textit{categories}. 
The latter is strictly related to the concept of \textit{richness}, which can be interpreted as the number of categories present in a system. 
The former instead is connected to the \textit{evenness} of a system, i.e. the distribution of elements across the categories. 
Richness and evenness are the two facets of what in the literature is called \textit{dual-concept diversity} (\cite{McDonald2003}). 
Along with them, \textit{disparity} is a third dimension of diversity, describing the difference between categories (\cite{Stirling2007}). 
%

Nonetheless, even if Stirling's definition can be easily generalizable to different contexts, it is fundamental to notice that approaching diversity several interpretations can be adopted, according to the context of use. 
Indeed, to completely abstract from the social context in which a technology is implanted when modelling diversity can be misleading, as \cite{Selbst2018} discuss. 
Even if the authors focus on the concept of fairness, likewise the issues identified can be found while treating diversity, considering the several common aspects between these two values, as pointed out by \cite{Celis2016}. 
Similarly,  \cite{Mitchell2020} discuss the link between fairness and diversity, and emphasize the differences between the idea of \textit{heterogeneity} related to the variety of any attribute within a set of instances, in comparison to \textit{diversity} intended as variety with respect to sociopolitical power differentials, such as gender and race.
As \cite{Drosou2017} affirm analyzing diversity in Big Data applications, diversity can hardly be universally defined in a unique way. 


The issues which arise in the conceptualization of diversity are reflected when attempts are made for establishing a universal formula to audit the diversity of a system. 
However, in several fields different needs have led to the formulation of different measurements, nowadays still in use and effective. 
Following, we refer to a diversity index as a measure able to quantify the relationship between elements distributed in categories of a system.

Two diversity indexes still being widely used have been proposed at the end of the 1940s: the commonly called \textit{Shannon index} (H')  (\cite{shannon1948mathematical}), and the \textit{Simpson index} (D) (\cite{simpson1949measurement}). 
Even if originally from two different fields, namely Information Theory and Ecology, both are based on the idea of choice and uncertainty.
Indeed, Shannon defines his formula wondering what measure would be suitable to describe the degree of uncertainty involved in choosing at random one event within a set of events. 
Similarly, Simpson formulated his index measuring the probability of choosing randomly two individuals from the same group within a population. 

The main limitation of these indexes is their focus on the analysis of the frequency of the elements, leaving aside any semantic information. 
\cite{BarHillel1953} discuss this limit, considering also the meaning of symbols in contrast to the frequentist approach. 
This semantic gap of diversity measurements can be partly solved by the introduction of a third dimension of diversity, \textit{disparity}, which joins variety and balance by creating a more solid framework for diversity analysis. 
%
%
This dimension is present in the Rao-Stirling diversity index: 
\begin{equation}
\Delta = \sum_{i,j \hspace{0,1cm}i\ne j}(d_{ij})^\alpha (p_i \cdot p_j)^\beta
\end{equation}

where $d_{ij}$ indicates the disparity between elements i and j, while $p_i$ and $p_j$ the proportional representations of those elements. %
This index initially proposed by \cite{Rao1982}, and revisited by \cite{Stirling2007}, is often considered while analyzing research interdisciplinarity in Scientometrics studies, even if its validity is still being discussed, as recently done by \cite{Leydesdorff2019}. 

In the next sections, we focus separately on the indexes we will use for our diversity analysis.

\subsection{Shannon Index}
\begin{equation}
    H' = - \sum_{i=1}^{S} p_i \ln{p_i}
\end{equation}

Consider that $p=n/N$ is the proportion of individuals of one particular species found $n$ divided by the total number of individuals found $N$, and $S$ is the number of species.
 
The Shannon index takes values between 1.5 and 3.5 in most ecological studies, and the index is rarely greater than 4. This measure increases as both the richness and the evenness of the community increase. The fact that the index incorporates both components of biodiversity can be seen as both a strength and a weakness. It is a strength because it provides a simple, synthetic summary, but it is a weakness because it makes it difficult to compare communities that differ greatly in richness.

\subsubsection{Pielou Index}
The Shannon evenness, discarding the richness, can be computed by means of the Pielou diversity index (\cite{pielou1966measurement}):

\begin{equation} \label{pielou}
    J' = \frac{H'}{H'_{max}}
\end{equation}

$H'$ is the Shannon diversity index and $H'_{max}$ is the maximum possible value of $H'$ (if every species was equally likely): 
\begin{equation}
    H'_{max} = - \sum_{i=1}^{S} \frac{1}{S}\ln{\frac{1}{S}} = \ln{S}
\end{equation}

$J'$ is constrained between 0 and 1, meaning 1 the highest evenness.

\subsection{Simpson Index}
The Simpson diversity index is a dominance index because it gives more weight to common or dominant species. In this case, a few rare species with only a few representatives will not affect the diversity. Since $D$ takes values between 0 and 1 and approaches 1 in the limit of a mono-culture, $1-D$ provides an intuitive proportional measure of diversity that is much less sensitive to species richness. Thus, Simpson's index is usually reported as its complement $1-D$.

\begin{equation}
    D = \frac {\sum_{S} n(n-1)}{N(N-1)}
\end{equation}

\section{Diversity indexes for AI conferences}
This work proposes several diversity indicators to measure gender, geographical and business diversity in top Artificial Intelligence conferences. Gender diversity is the main focus of programs such as Women in ML\footnote{\url{https://wimlworkshop.org/}}; Geographic diversity is linked to the presence of different countries and cultures in AI research. Finally, academia vs industry provides a way to assess the type of institutions contributing to AI research. We think these are three key socio-economic aspects of AI communities. All our indicators base their formulation in the biodiversity indexes described in the previous sections.

\subsection{Gender Diversity Index (GDI)}
We consider $S$ different species in the gender dimension: "male", "female" and any gender identity beyond the binary framework. We should distinguish between two possible cases: 
\begin{itemize}
    \item Only three species collected ("male", "female" and "other"). In this case, richness is not so relevant, while evenness gains more importance; therefore, we can discard the Simpson index and compute the Shannon evenness (we discard richness) by means of the Pielou diversity index.
    \item More species (gender identities) collected: we can apply the Shannon index in order to measure the evenness together with the richness. 
\end{itemize}

To compute the Diversity Index, we consider three different communities, as they represent complementary contributions to the scientific event: keynotes ($k$), authors ($a$) and organisers ($o$). Our final GDI performs a weighted average among the Pielou index in each community:

\begin{equation} \label{gdi}
    GDI = w_k J'_k + w_a J'_a + w_o J'_o
\end{equation}

As a default, we provide the same weight to keynotes, authors and organisers, although this can be configured to give more relevance to certain groups:

\begin{equation} \label{weights}
W = [w_k, w_a, w_o] = [\frac{1}{3}, \frac{1}{3}, \frac{1}{3}]
\end{equation}

\subsection{Geographical Diversity Index (GeoDI)}
In order to compute the Geographical Diversity Index we consider the same three communities: keynotes, authors and organisers. As we have multiple species (countries), we want to measure the richness together with the evenness, so we apply the weighted average of the Shannon Index community (this index may be greater than 1), using the weights $W$ defined in Equation \ref{weights}:

\begin{equation}
    GeoDI = w_k H'_k + w_a H'_a + w_o H'_o
\end{equation}

This index could also be computed by using the Simpson Index, but this would avoid the effect of very infrequent species (few people from some countries). 

\subsection{Business Diversity Index (BDI)}
The Business Diversity Index aims to compute the diversity of a conference regarding the presence of industry, academia and research centres. Thus, we apply Equation \ref{pielou}, considering $S=3$ when computing $H'_{max}$ (similar to GDI considering 3 species). Weights $W$ are defined in Equation \ref{weights}:

\begin{equation}
    BDI = w_k J'_k + w_a J'_a + w_o J'_o
\end{equation}

\subsection{Conference Diversity Index (CDI)}
The general Diversity Index of a Conference (CDI) is computed by averaging GDI, GeoDI and BDI. 
The typical values for the Shannon index are generally between 1.5 and 3.5 in most ecological studies, being rarely greater than 4. Therefore, GeoDI needs to be normalized between $[0,1]$ before being combined with the other indexes, so we divide it by 3.5. See Table~\ref{indexes} that summarises all the proposed indexes. 

\begin{equation}
    CDI = \frac{GDI + \frac{GeoDI}{3.5} + BDI}{3}
\end{equation}

\begin{table*}[!htbp]
 \caption{Diversity Indexes.}
  \centering
  \begin{tabular}{llll}
    Index     & Notation     & Based on & Range \\
    \cline{1-4}
    Gender Diversity Index & GDI & Pielou/Shannon Index & $[0,1]/[0,4]$ \\
    Geographical Diversity Index & GeoDI & Shannon Index & $[0,4]$ \\
    Business Diversity Index & BDI & Pielou Index & $[0,1]$ \\
    \textbf{Conference Diversity Index} & CDI & - & $[0,1]$ \\
  \end{tabular}
  \label{indexes}
\end{table*}

\section{Indexes evaluation}
In this section we describe the procedures of handling the data in order to evaluate the suitability of the proposed indexes to represent the diversity of major AI events. 
\subsection{Dataset}
The data publicly available from AI conferences is restricted to the name, surname and affiliation of authors, keynotes and organisers. This leads to some limitations when computing the proposed indicators. For instance, no data is provided about gender, so it needs to be inferred based on the name and surname, which introduces some errors and oversimplification to binary labels. Another limitation affects the way in which we compute the geographical diversity index. Having information just about the affiliation and not about the nationality, makes ethnic-based analysis extremely difficult to be performed. However, these limitations can be solved if the conferences' organisers collect more data at registration time, for instance. We are aware that some of this data might be personal and sensitive information but it can be beneficial to extract this kind of statistical analysis, which should ensure strict privacy and governance rules.

In order to compute the diversity indexes, we need to measure $p=n/N$ (i.e.: the proportion of individuals of one particular species found $n$ divided by the total number of individuals found $N$), and $S$ (i.e.: the number of species). For this purpose, we collected the names and affiliations of keynotes, organisers and authors (of a random sample comprising 10\% of the papers) of four consecutive years of NeurIPS\footnote{NeurIPS: Conference on Neural Information Processing Systems}, RecSys\footnote{RecSys: The ACM Recommender Systems conference} and ICML\footnote{ICML: International Conference on Machine Learning}. The size of each sample is listed in Table~\ref{conferences}. This data was gathered on several hackfests using a collaborative web application\footnote{\url{https://divinai.org}} designed for this purpose, that we also used to engage with AI students and the research community (e.g. AAAI conference) and outreach on the relevance of the topic and the need for community efforts. All the project data and material is available openly so it can be reproduced and extended to other conferences. Note that our indicators are exclusively based on public-domain information available at conference proceedings.

\begin{table*}[!htbp]
 \caption{Analysed conferences and size of the collected samples per year. Note that the sample size includes authors, keynotes and organisers.}
  \centering

  \begin{tabular}{ll}
    Conference Acronym  &  Year (Sample Size) \\
    \cline{1-2}
    NeurIPS & 2017 (343), 2018 (215), 2019 (549), 2020 (851)  \\
    RecSys & 2017 (41), 2018 (68), 2019 (69), 2020 (70)  \\
    ICML & 2017 (137), 2018 (264), 2019 (358), 2020 (450)  \\
  \end{tabular}
  \label{conferences}
\end{table*}

\subsubsection{Computing GDI}
When computing the Gender Diversity Index, our dataset does not provide gender identity information, so we infer the gender based on the given first name and surname (in some cases, we made use of the NamSor gender classifier library\footnote{\url{https://v2.namsor.com/}}). 

Due to this limitation of the publicly available datasets for identifying more gender options, we got $S=2$ different species: "male" and "female" and we used the Pielou diversity index ([0,1]). As we mentioned before, this fact can be overcame if the dataset includes more gender options (for instance, if this data is collected by the organisers of a conference, using information provided during registration).

\subsubsection{Computing GeoDI}
In order to measure the Geographical Diversity Index, the available information in our dataset is just the country of the affiliation. This means that we might not be considering the nationality, but the current location of each individual. This limitation could be avoided by, again, asking for the nationality in the registration form and building the dataset with this information. 

\subsubsection{Computing BDI}
The Business Diversity Index aims to compute the diversity of a conference regarding the presence of industry, academia and research centres. Once again, the affiliation gives us this information, although in some cases some specific web search was needed to label the dataset. For authors with double affiliation, we consider the first one for the labeling process. In this case, we set $S=3$.

\begin{table*}[htbp]
 \caption{Gender Diversity Index (GDI), with the percentage of male and female among authors (from a random sample of 10\% of the papers), keynotes and organisers.}.
  \centering
  \begin{tabular}{lrrrrrrrc}

     Conference & \multicolumn{3}{c} {\%Female} && \multicolumn{3}{c}{\%Male} & \textbf{GDI} \\
    \cline{1-9}
    & Auth & Key & Org && Auth & Key & Org & \\

    NeurIPS 2020 & 20.01 & 42.90 & 47.10 && 79.09 & 57.10 & 52.90 & \textbf{0.90}\\
    NeurIPS 2019 & 16.60 & 42.90 & 51.90 && 83.40 & 57.10 & 48.10 & \textbf{0.89}\\
    NeurIPS 2018 & 7.10 & 42.90 & 20.90 && 92.90 & 57.10 & 79.10 & \textbf{0.70}\\
    NeurIPS 2017 & 9.45 & 42.90 & 21.30 && 90.05 & 57.10 & 78.70 & \textbf{0.73}\\
    RecSys 2020 & 8.46 & 33.30 & 23.70 && 91.50 & 66.70 & 76.30 & \textbf{0.69}\\
    RecSys 2019 & 12.70 & 100 & 23.10 && 87.30 & 0 & 76.90 & \textbf{0.42}\\
    RecSys 2018 & 14.30 & 66.70 & 30.40 && 85.70 & 33.30 & 69.60 & \textbf{0.80}\\
    RecSys 2017 & 15.30 & 0 & 13.60 && 84.70 & 100 & 86.40 & \textbf{0.35}\\
    ICML 2020 & 15.50 & 33.30 & 37.90 && 84.50 & 66.70 & 62.10 & \textbf{0.84} \\
    ICML 2019 & 11.40 & 66.70 & 38.10 && 88.60 & 33.30 & 61.90 & \textbf{0.63} \\
    ICML 2018 & 9.80 & 50.00 & 28.90 && 90.20 & 50.00 & 71.10 & \textbf{0.78} \\
    ICML 2017 & 7.70 & 50.00 & 29.40 && 92.30 & 50.00 & 70.60 & \textbf{0.76}\\
  \end{tabular}
  \label{GDI}
\end{table*}

\begin{table*}[htbp]
 \caption{Geographical Diversity Index (GeoDI), with the number of developing countries and continents represented (we only collected the authors of the 10\% of the papers).}
  \centering
  \resizebox{\textwidth}{!}{
  \begin{tabular}{lrrrcccccc}
     Conference & \multicolumn{3}{c}{\# Countries} && \multicolumn{3}{c}{\# Continents} & \textbf{GeoDI/3.5} & \textbf{GeoDI\_continents}\\
    \cline{1-10}
   & Auth & Key & Org && Auth & Key & Org & \\
    NeurIPS 2020 & 28 & 5 & 9 && 5 & 3 & 5 & \textbf{0.50} & \textbf{0.55}\\
    NeurIPS 2019 & 5  & 2 & 8  && 3 & 1 & 4 & \textbf{0.19} & \textbf{0.16}\\
    NeurIPS 2018 & 15  & 3  & 11  && 4 & 2 & 3 & \textbf{0.36} & \textbf{0.34}\\
    NeurIPS 2017 & 19  & 3  & 10  && 4 & 2 & 3 & \textbf{0.36} & \textbf{0.36}\\
    RecSys 2020 & 9  & 2  & 16  && 4 & 2 & 4 & \textbf{0.48} & \textbf{0.51}\\
    RecSys 2019 & 5  & 2  & 8  && 3 & 1 & 3 & \textbf{0.36} & \textbf{0.33}\\
    RecSys 2018 & 9  & 2  & 9  && 3 & 1 & 3 &\textbf{0.42} & \textbf{0.31}\\
    RecSys 2017 & 5  & 2  & 10  && 3 & 2 & 4 & \textbf{0.37} & \textbf{0.43}\\
    ICML 2020 & 25  & 2  & 5  && 5 & 2 & 3 & \textbf{0.33}  & \textbf{0.38}\\
    ICML 2019 & 20  & 2  & 3  && 3 & 2 & 2 & \textbf{0.30} & \textbf{0.35}\\
    ICML 2018 & 14  & 2  & 7  && 4 & 2 & 2 & \textbf{0.34} & \textbf{0.37}\\
    ICML 2017 & 13  & 3  & 6  && 3 & 2 & 4 & \textbf{0.41} & \textbf{0.46}\\
  \end{tabular}}
  \label{GeoDI}
\end{table*}

\begin{table*}[htbp]
 \caption{Business Diversity Index (BDI), presenting as well the percentage of authors (from a random sample of 10\% of the papers), keynotes and organisers belonging to Academia, Industry or Research Centres.}
  \centering
  \resizebox{\textwidth}{!}{\begin{tabular}{lrrrrrrrrrrrc}
      Conference & \multicolumn{3}{c} {\%Academia} && \multicolumn{3}{c}{\%Industry} && \multicolumn{3}{c}{\%Research Centre} & \textbf{BDI} \\
    \cline{1-13}
    & Auth & Key & Org && Auth & Key & Org && Auth & Key & Org & \\
    NeurIPS 2020 & 52.60 & 57.10 & 41.20  && 31.70  & 14.30 & 47.10 && 15.70 & 28.60 & 11.80 & \textbf{0.89}\\
    NeurIPS 2019 & 49.10 & 85.70 & 44.40  && 39.50  & 14.30 & 40.07 && 11.40 & 0 & 14.80 & \textbf{0.72}\\
    NeurIPS 2018 & 72.30 & 57.10 & 59.70  && 9.22  & 42.90 & 31.30 && 18.40 & 0 & 8.96 & \textbf{0.71}\\
    NeurIPS 2017 & 73.10 & 57.10 & 63.90  && 22.20 & 42.90 & 29.50 && 4.73 & 0 & 6.56 & \textbf{0.67}\\
    RecSys 2020   & 69.00 & 66.70 & 78.90 && 27.60 & 33.30  & 18.40 && 3.40 & 0 & 2.60 & \textbf{0.59}\\
    RecSys 2019   & 40.00 & 100.00 & 69.20 && 55.40 & 0  & 23.10 && 4.62 & 0 & 7.69 & \textbf{0.49}\\
    RecSys 2018   & 23.80 & 33.30 & 56.50 && 47.60 & 66.70  & 39.10  && 28.60 & 0 & 4.35 & \textbf{0.78}\\
    RecSys 2017   & 70.80 & 50.00 & 81.80  && 28.30 & 50.00 & 13.60 && 0.89 & 0 & 4.55 & \textbf{0.58} \\
    ICML 2020    & 48.70 & 66.70 & 58.60 && 39.20 & 33.30 & 20.70 && 
    12.10 & 0 & 20.70 & \textbf{0.68} \\
    ICML 2019    & 43.10 & 66.70 & 76.20 && 42.50 & 33.30 & 19.00 && 14.40 & 0 & 4.76 & \textbf{0.70} \\
    ICML 2018    & 66.70 & 100.00 & 77.80 && 27.10 & 0 & 17.80 && 6.19 & 0 & 4.44 & \textbf{0.44} \\
    ICML 2017    & 51.80 & 50.00 & 88.20 && 34.20 & 25.00 & 11.80 && 14.00 & 25.00 & 0 & \textbf{0.72} \\
  \end{tabular}}
  
  \label{BDI}
\end{table*}

\subsection{Results}
In this section, we analyse the diversity indexes computed for the set of selected conferences. We structure the analysis in four different parts, corresponding to the four diversity indexes: GDI, GeoDI, BDI and the general CDI.

Table~\ref{GDI} reports the percentage of male and female among authors, keynotes and organizers. In general, the values obtained for GDI are quite high (over 0.50), except for  $GDI(RecSys 2017)=0.35$ and $GDI(RecSys 2019)=0.42$), as it is penalising the presence not just the low presence of female authors and organisers but also the presence of just one gender among the keynotes (in 2017 all keynotes were "male" and in 2019 all keynotes were "female"). If we focus on the rest of the conferences, we observe that there are efforts in the scientific community to balance gender among the keynotes. However, organisers and authors are mostly "male" and women do organize more than authorise. Those conferences balancing also the gender among organisers (NeurIPS 2019 and NeurIPS 2020) got the highest GDI (0.89 and 0.9 respectively). 

The analysis regarding the Geographical Diversity Index is summarised in Table~\ref{GeoDI}. As we mentioned before, we normalised it to the range [0,1] to make it comparable with the other indexes. As an index based just on the number of countries might hide a lack of diversity regarding, for instance, the presence of researchers from least developed countries, we also computed the number of developing countries present (following the United Nations classification\footnote{\url{https://unctad.org/topic/least-developed-countries/list}}). We couldn't find any representation of any of the countries included in this list containing 46 countries. Thus, we also grouped the affiliations data in order to report the presence of continents and explore the variability of the index in considering these major geographical divisions. The $GeoDI_{continents}$ index is computed following the Pielou Index formula (see Equation~\ref{pielou}) and is also listed in Table~\ref{GeoDI}. It belongs to the range [0,1], so normalization is not needed.

We can see that, in general, the indexes computed for the continents are very similar to those related to the countries, and they have a value below 0.5. We consider 7 continents (Africa, Antarctica, Asia, Europe, North America, South America and Oceania), in order to avoid hiding lower representation of Latin American countries. In most of the conferences explored, there are few species (usually 3 -North America, Europe and Asia- and rarely 4 -including Oceania-). Moreover, these countries are not equally represented: most of the keynotes come from North America or Europe and just 5 researchers from Africa were found among authors and organisers. We would also like to note the importance of the conference location for promoting the presence of minorities. We highlight the case of RecSys 2020, located in Brazil, with a high representation of organisers (13 out of 38) and even one keynote (out of 3) from South America, a continent with almost no representation in the rest of the events.

Table~\ref{BDI} reports the Business Diversity Index for the studied conferences. Again, NeurIPS 2020 presents the higher diversity index (0.89), as it has representation from all sectors. On the other side, the lowest indexes are reached by ICML 2018 (0.44) and RecSys 2019 (0.49) as all the keynotes belong to Academia. In general, most of the conferences are academic (having low representation from industry or research centres). In the case of RecSys 2018 (0.78), this good index is due the great balance of the different species even if there is not keynotes belonging to a Research Centre. 

\begin{figure}[ht]
 \centering
  \includegraphics[width=\columnwidth]{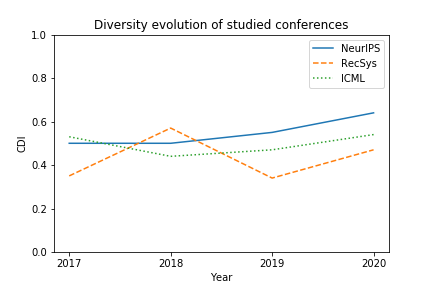}
\caption{Conference Diversity Index (CDI) evolution.}
\label{fig:CDI}
\end{figure}

We observe that the diversity indexes can provide more information at a glance than the other measures reported (percentages, number of countries...). In fact, the general Conference Diversity Index (CDI) aims to summarize, using just one value, the gender, geographical and business diversity of a conference. This index also provides a very useful measure to monitor the diversity evolution of a conference, and easily compare it with other conferences of the same topic. Figure~\ref{fig:CDI} shows how the different conferences evolve in terms of the Conference Diversity Index. Values of CDI over 0.5 means that the conference achieves acceptable diversity indexes. For a more detailed information about this indicator, we should explore each index separately. 

\section{Conclusions}

This work aims to raise awareness about the lack of diversity in Artificial Intelligence, by defining a set of indicators that measure the gender, geographical and business diversity of 
international AI conferences. We have explored a set of recent top international AI conferences in order to compute their related indexes and compare them in terms of diversity. Our results indicate a huge gender unbalance among authors, a lack of geographical diversity (with no representation of least developed countries and very low representation of African countries). However, we could show evidence of the recent efforts done in promoting minorities among keynote speakers, reaching gender balance in several conferences. 

We should note that these preliminary conclusions are limited by the conference sample, the considered gender proxy (simplified to three classes and labeled based on name, with potential errors), and the fact that the geographical proxy relates to affiliation and not nationality. We will consider these aspects in future work and developments.

However, we think that our proposed formulation for measuring diversity can be extremely useful for conference organisers, as they can have access to more detailed data that can shed light on the lack of diversity from different perspectives. With this information, organisers should focus on those indexes under 0.5 and try to increase them by launching specific actions for promoting the identified minorities: scholarships or discounts for underrepresented collectives, celebrating conferences in locations with lower representation, more diversity among keynote speakers and organisers, etc. We would like to note that the indexes proposed in this work can easily be applied to conferences of different topics. 

As future work, we would like to explore the viability of studying the ethnic diversity based on the analysis of the names and surnames. Also, we will study how to improve the indexes definition, focusing on how to normalize them in a more suitable way, as the GeoDI index might be a bit penalised with the current normalisation. 
Finally, we would like to incorporate more conferences to be able to perform some long-term analysis. 

\acks{This work has been partially supported by the HUMAINT programme (Human Behaviour and Machine Intelligence), Centre for Advanced Studies, Joint Research Centre, European Commission (EC) and the Spanish Ministry of Economy and Competitiveness under the Maria de Maeztu Units of Excellence Programme (MDM-2015-0502). Lorenzo Porcaro acknowledges financial support from the EC under the TROMPA project (H2020 770376).}

\bibliography{jmlr-sample}
\end{document}